\documentclass[aps,prb,twocolumn,groupedaddress,showpacs]{revtex4}
\usepackage{graphicx}
\usepackage{dcolumn}
\usepackage{bm}
\usepackage{amsmath,amsfonts,amssymb,amsthm,verbatim}
\usepackage[ansinew]{inputenc}
\usepackage[usenames]{pstricks}
\usepackage{color}
\usepackage{epsfig}

\newcommand{\be}{\begin{equation}}
\newcommand{\ee}{\end{equation}}
\newcommand{\bea}{\begin{eqnarray}}
\newcommand{\eea}{\end{eqnarray}}

\newcommand{\bi}{\begin{itemize}}
\newcommand{\ei}{\end{itemize}}
\newcommand{\bc}{\begin{center}}
\newcommand{\ec}{\end{center}}

\begin{document}

\title{$\text{Robustness of symmetry-protected topological states against time-periodic perturbations}$}

\author{Oleksandr Balabanov}
\affiliation{Department of Physics, University of Gothenburg, SE 412 96 Gothenburg, Sweden}

\author{Henrik Johannesson}
\affiliation{Department of Physics, University of Gothenburg, SE 412 96 Gothenburg, Sweden}

\begin{abstract}

The existence of gapless boundary states is a key attribute of any topological insulator. Topological band theory predicts that these states are robust against static perturbations that preserve the relevant symmetries. In this article, using Floquet theory, we examine how chiral symmetry-protection extends also to states subject to time-periodic perturbations $-$ in one-dimensional Floquet topological insulators as well as in ordinary one-dimensional time-independent topological insulators. It is found that, in the case of the latter, the edge modes are resistant to a much larger class of time-periodic symmetry-preserving perturbations than in Floquet topological insulators. Notably, boundary states in chiral time-independent topological insulators also exhibit an unexpected resilience against a certain type of symmetry-breaking time-periodic perturbations. We argue that this is a generic property for topological phases protected by chiral symmetry. Implications for experiments are discussed.

\end{abstract}

\maketitle

\section{Introduction} A hallmark of a symmetry-protected topological (SPT) phase of matter $-$ topological insulators and topological superconductors being well-known examples~$\text{\cite{Hasan, Qi}}$ $-$ is the presence of gapless boundary states~\cite{Ryu}.  
The very existence of these states is a consequence of the nontrivial topology of the bulk band structure (``bulk-boundary correspondence"), with the symmetry protection ensuring their robustness against static gap-preserving perturbations as long as the relevant symmetries remain unbroken \cite{Volovik, Kitaev, Essin}. As is well known, this robustness, along with other unique properties of the boundary states, has raised the prospects for exploiting SPT phases for future technologies $-$ from applications in spintronics \cite{Pesin} to topological quantum computation \cite{Stanescu}.

How does the symmetry protection play out when the boundary states are subject to {\em time-dependent} perturbations? While a comprehensive answer has to await further advances in the theory of SPT phases, the case of {\it time-periodic} perturbations can be addressed efficiently by using Floquet theory \cite{Shirley, Sambe}. In this work we exploit this advantage to answer the question to what extent the midgap boundary states in a SPT phase are robust against a time-periodic disordering perturbation. While such perturbations may not occur naturally in physical systems, they have recently been realized in highly controlled experiments with cold atomic \cite{CA_driven_disorder} and optical \cite{PC_driven_disorder} setups. As such, their study could open a new inroad to explore the physics of SPT phases.

We focus on two typical brands of one-dimensional (1D) SPT phases $-$ Floquet topological insulators \cite{Kitagawa, Lindner} and ordinary time-independent insulators~\cite{Hasan, Qi} $-$ with boundary states protected by chiral symmetry \cite{Ryu}. The first type $-$ the Floquet insulator $-$ is obtained by driving a system with a time-periodic field, 
resulting in symmetry-protected boundary states in nonequilibrium \cite{Kitagawa, Lindner, Rudner, Reynoso, Liu, Leon, Potirniche}. The protection of the states comes about as a feature of the time-evolution operator, implying that robustness against added time-periodic perturbations can be investigated within the same Floquet formalism which describes the driving of the bulk. As noted by Asb\'{o}th {\em et al.}~\cite{Asboth}, the robustness of boundary states in a chiral Floquet system driven by two time-periodic fields of the same frequency depends critically on the relative phase of the driving. By a systematic study we here extend this picture to the case where one field drives the bulk, with the other acting as a time-periodic disordering perturbation. As one would expect, we find that boundary states remain robust to time-periodic perturbations that preserve chiral symmetry. In its turn, the very existence of chiral symmetry in an unperturbed driven system depends crucially on the phase of the bulk driving. It follows that the states remain protected only when the phase of the time-periodic disordering perturbation is properly tuned to the driving in the bulk. This is explicitly shown analytically and confirmed numerically.

Turning to the time-independent topological insulators, we can again use Floquet theory to study the effect of time-periodic perturbations. This is so, since any time-independent Hamiltonian is trivially periodic in time. Our analysis and findings here can be summarized as follows: First, we make explicit how the freedom in choosing starting point in the stroboscopic Floquet time evolution implies that chiral time-independent systems actually possess an infinite number of chiral symmetries. As a consequence, the edge states are robust against a much larger class of symmetry-preserving time-periodic perturbations compared with those of a Floquet topological insulator. Second, we establish a class of symmetry-breaking time-periodic perturbations for which the boundary states display an unexpected resilience. A detailed analysis reveals how this property is manifested in Floquet perturbation theory: The effect coming from this class of perturbations gets suppressed by the very structure of the unperturbed chiral-symmetric spectrum, implying that its expected leading-order contribution vanishes identically. 
Such protection is a very interesting feature because it hints that, even when the chiral symmetry is broken, a residue of it can still have an effect on the system's behavior.

For simplicity, our analysis proceeds by way of example, with the Su-Schrieffer-Heeger (SSH) model~\cite{SSH} as a case study. To cover the two classes of topological insulators $-$ Floquet and time-independent ones $-$ we consider a periodically driven version of the SSH model as well as the original time-independent variety. Our choice is motivated by the fact that the SSH model serves as a prototype for band insulators exhibiting topological phases \cite{SSH2, SSH3, SSH4, SSH5}. We should point out, however, that our analysis can be carried over to any 1D chirally symmetric topological phase (symmetry classes AIII, BDI, DIII, and CII in the Altland-Zirnbauer classification \cite{Ryu2}).

The paper is planned as follows: In the next section, after a brief introduction to Floquet formalism, we present the harmonically driven SSH model and describe its topological properties. We then discuss the topological protection of the boundary states and explicitly identify the types of time-periodic disorder in the presence of which the states remain robust. 
In Sec.~III we turn to the undriven (time-independent) SSH model and describe first the symmetries that the system possesses within Floquet theory. It is then argued that the boundary states are robust to a much broader class of time-periodic perturbations than in the driven case and this is verified numerically. Next, we present our perturbative analysis revealing the enhanced resilience of the boundary states for a certain type of symmetry-breaking time-periodic perturbations. This is followed by a qualitative argument why this property may hold also outside the perturbative regime, supplemented by supporting data from numerical computations. In the same section we discuss time-periodic disorder in the chemical potential $-$ an important type of perturbations for making contact with experiments. 
Finally, we comment on the feasibility to test our predictions in an experiment using optically trapped cold atoms. A brief summary and outlook is given in~Sec.~IV.

\section{Harmonically driven SSH model}

\subsection{Floquet formalism}

To set the stage, let us recall that the time evolution of any quantum system driven by time-periodic fields can be described using Floquet theory \cite{Shirley, Sambe}. Within this formalism an equivalent of energies, so-called {\em quasienergies}, can be defined and one may consider the band structure of the system in terms of its quasienergy spectrum. Specifically, the Schr\"{o}dinger equation with a time-periodic Hamiltonian $H(t) = H(t+T)$ has a complete set of solutions $|\psi_n(t)\rangle = \exp(-i\varepsilon_n t) |u_n(t)\rangle$, commonly called {\em steady states}, where $\varepsilon_n$~denotes the quasienergies and $|u_n(t)\rangle$ = $|u_n(t+T)\rangle$ for all times $t$ (with $\hbar\!\equiv \!1$). The quasienergies, defined modulo~$2 \pi /T$,  appear in the dynamical phase acquired by the steady states, and in this sense they are similar to ordinary energies. The quasienergy spectrum can be found by using  the fact that the states $|u_n(t)\rangle$ are eigenstates of the evolution operator $U (t, T+t)$ associated with the eigenvalues $\exp(-i\varepsilon_n T)$. To find the band structure it is thus sufficient to diagonalize $U (t, T+t)$ for some conveniently chosen fixed time~$t$.  Alternatively, one can Fourier transform the Schr\"{o}dinger equation and perform the calculations in the frequency domain~\cite{Shirley, Sambe}.

\subsection{Topological characteristics of the model}

The SSH model consists of spinless fermions hopping on a 1D staggered lattice~\cite{SSH}. Here we assume that the hopping amplitudes have both static and time-dependent harmonically modulated components as shown in Fig.~1. Within a tight-binding approximation the Hamiltonian with vanishing chemical potential can be written as~\cite{Lago}

\begin{align}
\begin{split}
H(t)  = & - \sum_{j} \left( \gamma_1 c^{\dagger}_{A,j} c^{\phantom\dagger}_{B,j} + \gamma_2 c^{\dagger}_{B,j-1} c^{\phantom\dagger}_{A,j} + \text{H.c.} \right ) \\ 
& + \sum_{j} \left( v(t) c^\dagger_{A,j} c^{\phantom\dagger}_{B,j} - v(t) c^\dagger_{B,j-1} c^{\phantom\dagger}_{A,j} + \text{H.c.} \right ),
\end{split}
\label{eq:SSH_harmonic}
\end{align}
where $c^\dagger_{\sigma,j}$ and $c^{\phantom\dagger}_{\sigma,j}$ ($\sigma=A, B$) are creation and annihilation operators, $\gamma_1$ and $\gamma_2$ are the static intracell and intercell hopping amplitudes respectively, and $v(t)=2 V_{\text{ac}} \cos(\Omega t)$ is the harmonically modulated component of the hopping. 

\begin{figure} \centering
    \includegraphics[width=8.0cm,angle=0]{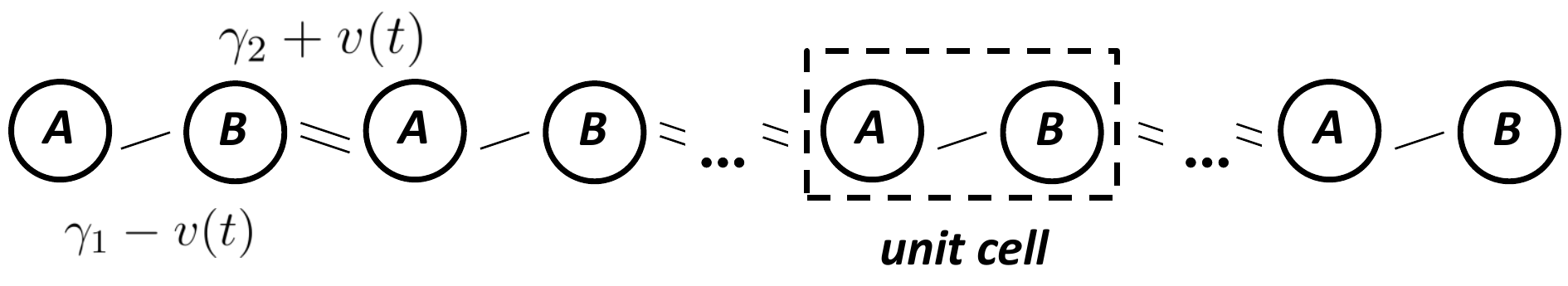}
    \caption{A schematic illustration of the harmonically driven SSH chain, with hopping amplitudes $\gamma_{1/2} \pm v(t)$, where $v(t) \sim \cos(\Omega t)$.}
     \label{fig_SSH}
\end{figure}

The undriven SSH Hamiltonian $H_0$ [defined by setting $v(t)=0$ in Eq. (1)] has chiral symmetry, implying the existence of a unitary operator $\Gamma$ such that $\Gamma H_0 \Gamma = - H_0$ where $\Gamma c_{A,j} \Gamma = c_{A,j}$ and $\Gamma c_{B,j} \Gamma = - c_{B,j}$. The representation~[28]
\begin{equation}
\Gamma \equiv e^{i\pi \sum_j c^\dagger_{B,j} c^{\phantom\dagger}_{B,j}}
\end{equation}
allows for an easy check of chiral symmetry in the second-quantized formalism, more convenient than if one were to directly use a representation of the second-quantized antiunitary chiral operator~[26]. Clearly, the chiral symmetry here reflects a sublattice symmetry, i.e., the property that the SSH Hamiltonian $H_0$ does not couple sites on the same sublattice. As it turns out, the harmonically driven SSH model has a chiral symmetry as well, but now given for the evolution operator. This can be established by first noticing that $H(t)$ in Eq. (\ref{eq:SSH_harmonic}) satisfies the relation

\begin{align}
\begin{split}
\Gamma H(t) \Gamma = - H(-t).
\end{split}
\label{eq:SSH_chiral_symmetry_t}
\end{align}
As shown in Appendix A, this property is sufficient for proving that the evolution operators $\text{$F \equiv U(0,T/2)$}$ and $\text{$G \equiv U(T/2, T)$}$ are related as $F = \Gamma G^\dagger \Gamma$. From this follows immediately that the periodically driven model has chiral symmetry, as expressed by $\Gamma U (0,T)\Gamma = U^{-1}(0,T)$~\cite{Asboth}. 

\begin{figure} \centering
    \includegraphics[width=7cm,angle=0]{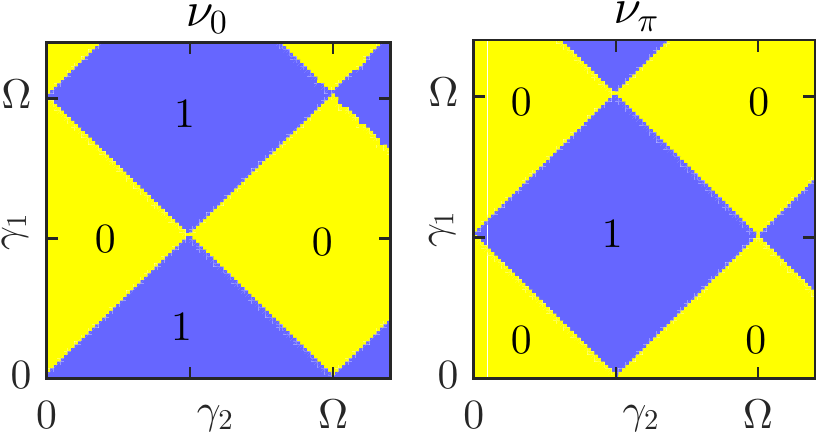}
    \caption{The topological invariants $\nu_0$ and $\nu_\pi$ calculated for $V_{\text{ac}}=0.2 \, \Omega$ and different values of 
    $\gamma_1$ and $\gamma_2$. 
    $\nu_{\pi/0} =1 (0)$ corresponds to a topologically nontrivial (trivial) winding number.}
     \label{fig1}
\end{figure}

With chiral symmetry in hand, we refer to a general result \cite{Asboth} to conclude that the topological phases of the harmonically driven SSH model can be characterized by two integer topological invariants, $v_0$ and $v_\pi$. These invariants count the number of SPT boundary states at each end of the chain, corresponding to quasienergies~$0$ and~$\Omega/2$ respectively. It has been suggested that both of them can be extracted from the operator $F$ defined above, after imposing spatially periodic boundary conditions \cite{Asboth, Asboth2}. For the present problem we have carried out the calculation numerically by discretizing the time-evolution into a large number of intervals and assuming that the Hamiltonian $H(t)$ is constant in each of them. The values of the topological invariants for different static hopping amplitudes are displayed in Fig.~\ref{fig1}, in excellent agreement with that obtained from an analysis of the Zak phase \cite{Lago}.

\subsection{Protection against symmetry-preserving boundary perturbations}

\begin{figure} \centering
    \includegraphics[width=8.0 cm, angle=0]{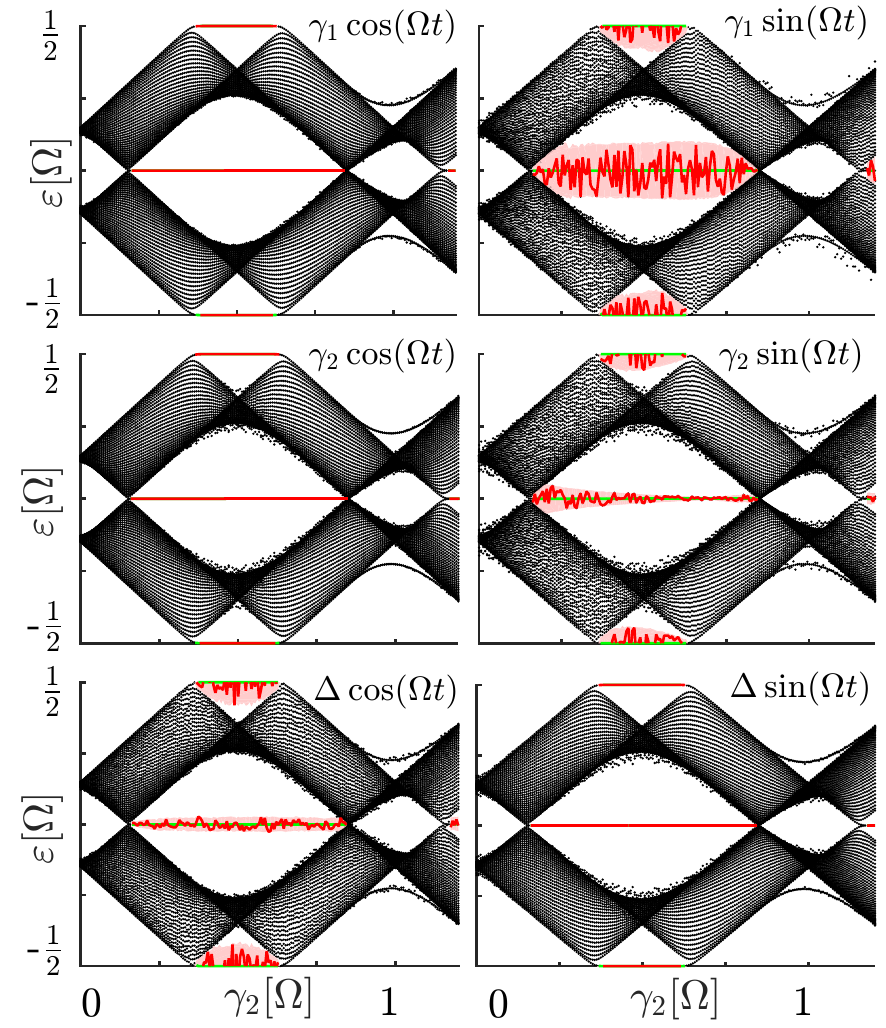}
    \caption{Quasienergy spectra of harmonically driven SSH chains subject to time-periodic boundary perturbations, here realized as a spatial disordering of the amplitudes for hopping $(\sim \gamma_1, \gamma_2)$ or of an added staggered chemical potential ($\sim \Delta$). The chains have 80 sites, with unperturbed $\gamma_1 = 0.15 \, \Omega$, $V_{\text{ac}}=0.2 \, \Omega$, and with $\gamma_2$ swapped from $0$ to $1.2 \, \Omega$. The disorder is added over 20 sites from one of the boundaries, with the disorder amplitudes varying randomly within the interval $[- 0.2 \, \Omega, 0.2 \, \Omega]$. Here we display levels for bulk states (black), perturbed (red), and unperturbed (green) edge states corresponding to single disorder realizations. For each type of perturbation 100 different disorder realizations were considered, with the midgap quasienergies always to be found confined to the corresponding pink regions. }
     \label{fig_SSH_time-dep}
\end{figure}

In exact analogy to the time-independent case, the symmetry protection of the topological invariants against time-periodic perturbations comes from the restriction that the chiral symmetry places on the quasienergy spectrum. For details, see  Appendix B. As a consequence, the boundary states in the thermodynamic limit of the SSH model, Eq. (\ref{eq:SSH_harmonic}), are expected to be robust against gap-preserving time-periodic perturbations $V(t)$ which do not violate the chiral relation $\Gamma V(t) \Gamma\!=\!-V(-t)$. This condition is satisfied for any site-dependent perturbation of the hopping amplitudes $\gamma_1$ and $\gamma_2$ that is {\em even} in time (which, trivially, includes static perturbations).  
In contrast to a disordering of the hopping amplitudes, a perturbation from an added time-periodic staggered chemical potential $\Delta$ (proportional to $c^\dagger_{A,j} c^{\phantom\dagger}_{A,j}- c^\dagger_{B,j} c^{\phantom\dagger}_{B,j}$) has to be {\em odd} in time in order to respect chiral symmetry~\cite{Asboth}. 
 
In  Fig.~3  we  numerically examine the robustness of the boundary states against various types of time-periodic perturbations, here added to one of the boundary regions (taken to extend over several sites near the left edge of the chain) as a spatial disordering of the amplitudes of the hopping $(\sim \gamma_1, \gamma_2)$ or the staggered chemical potential $(\sim \Delta)$. By confining the perturbation to a boundary region, we are ensured that the quasienergy bulk gaps stay open for {\it all} disorder realizations. In contrast, bulk perturbations may close the gap for certain realizations of large-amplitude disorder, removing the protection. The quasienergies were obtained by truncating the Hamiltonian in the frequency domain \cite{Sambe} and then diagonalizing it numerically. All perturbations were chosen to be harmonic in time, however, our approach can be generalized to any time-periodic perturbation. The numerical results validate our predictions above: The midgap levels at quasienergies 0 and $\Omega/2$ (in red color in Fig. 3), corresponding to the perturbed boundary states, are robust against perturbations of hopping amplitudes (staggered chemical potential) which are {\em even (odd)} in time, otherwise not. It is important to realize that the zero reference time gets fixed by us when we take the bulk driving to be proportional to $\cos(\Omega t)$. 
Thus, the phases of the allowed boundary perturbations for which the midgap states remain robust are determined by the phase of the bulk driving. We should also mention that the hopping disorders were taken to be complex in the computations, thus disabling particle-hole symmetry to protect the boundary states when the chiral symmetry is broken by a perturbation.

It is interesting to note the appearance of additional edge states in the topologically trivial regime of the harmonically driven SSH chains, with values of $\gamma_2$ near $\Omega$ (see Fig.~3). Similar states have been seen also in other 1D Floquet systems \cite{Saha}. These boundary modes are not expected to be of a topological origin. Still, such states are robust against weak perturbations because they are separated from the bulk modes by a finite gap. Intriguingly, the response of these states to various time-periodic perturbations seems to be correlated with the robustness of the topological edge states: In Fig.~3 we see that the corresponding quasienergy shifts are much more profound in the cases where chiral symmetry is broken by the perturbation. This feature warrants further study.

\section{Time-independent SSH model } 

\subsection{Symmetry preservation in the model} 
 
 Time-independent models can also be handled within the Floquet formalism because any static Hamiltonian is periodic in time for any frequency~$\Omega$. Thus, we may write the evolution operator of a time-independent model as $U(t_0, t_0 +T)$, where $t_0$ is a fixed reference time and $T$~is interpreted as a period in the Floquet formalism. By this simple change of perspective we can systematically explore the robustness of the boundary states against time-periodic perturbations. It should be stressed that once we enter Floquet theory the notion of energy is replaced by quasienergy and this must be carefully taken into account.  
 
Having expressed the evolution operator for a time-independent chirally symmetric model as $U(t_0, t_0 +T)$, it is essential to note that within the Floquet formalism the model actually supports an infinite number of chiral symmetries. This is so because $\Gamma U(t_0, t_0 +T) \Gamma \! = \! U^{-1}(t_0, t_0 +T)$ for any choice of reference time $t_0$. Since the effect of a time-periodic perturbation is independent of the choice of $t_0$, the perturbation has to break {\it all} these chiral symmetries in order to kick the quasienergies away from zero. Therefore, the symmetry-protected boundary modes in static chiral models are expected to be robust to a much broader class of time-periodic perturbations in comparison with the Floquet topological insulators discussed above. 
\begin{figure} \centering
    \includegraphics[width=8.0 cm, angle=0]{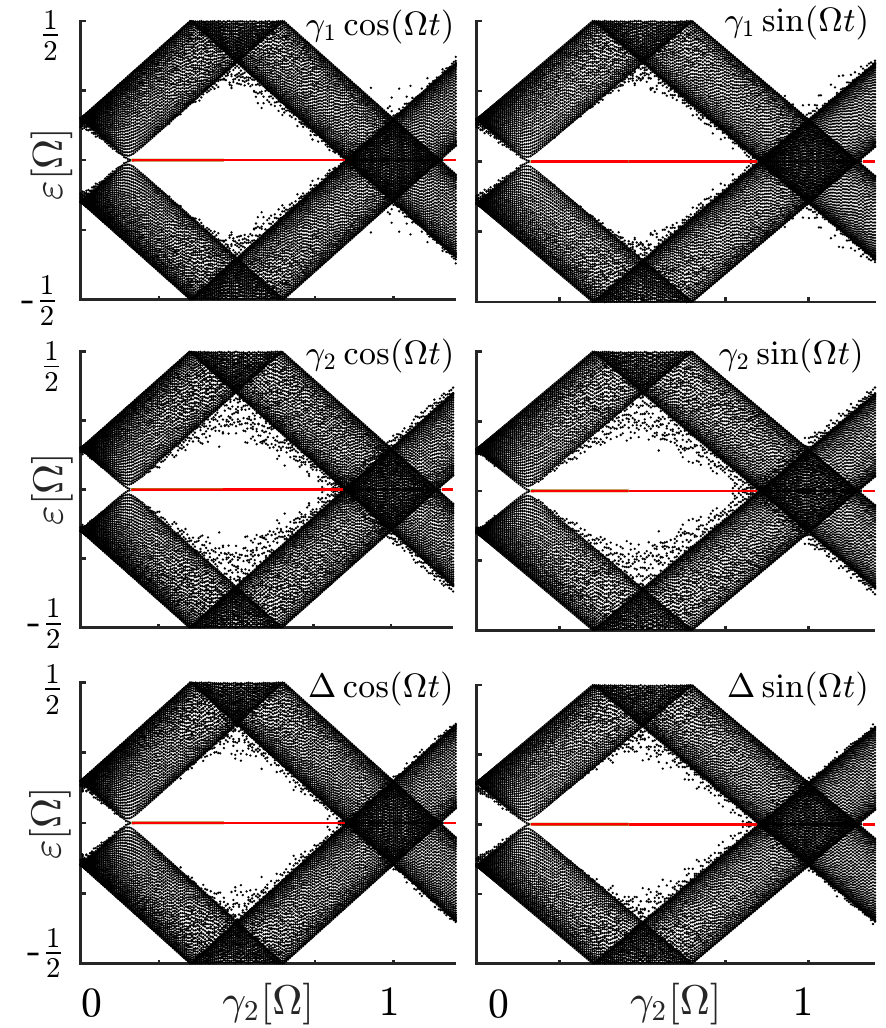}
    \caption{ Quasienergy spectra of time-independent SSH chains [$v(t) = 0$ in Eq. (1)] under influence of various time-periodic boundary perturbations. These perturbations were implemented as a spatial disordering of the hopping amplitudes $(\sim \gamma_1, \gamma_2)$  or of an added staggered chemical potential ($\sim \Delta$). The chains were taken with $\gamma_1 = 0.15 \, \Omega$ and $\gamma_2$ varying between $0$ and $1.2 \, \Omega$. Each chain consists of 80 sites and is disordered over the first 20 sites from one of the boundaries, with the corresponding disorder amplitudes chosen randomly within the interval $[- 0.2 \, \Omega, 0.2 \, \Omega]$. Here we illustrate quasienergies for bulk states in black and edge states in red, respectively (the quasienergies of the perturbed and unperturbed edge modes perfectly match in this case).
    }
     \label{fig_SSH_time-dep2}
\end{figure}

In Fig.~4 we show numerical data for undriven SSH chains [$v(t)\!=\!0$ in Eq.~(1)] subject to the same time-periodic disorders that we considered in Fig.~3 for the harmonically-driven case. The quasienergies corresponding to the symmetry-protected states are seen to be completely unaffected by the harmonic single-parameter disorders, independently of the phase of the perturbative driving. This is in full agreement with the discussion above because, in each of these cases, the chiral symmetry is preserved for some $t_0$. We briefly note that the band structure now supports only a single gap, with all dynamical gaps being closed because of the absence of bulk driving. Also, the edge states for $\gamma_2$ near $\Omega$ are not present anymore.

\subsection{Resilience of the boundary states against symmetry-breaking perturbations}

In general, topological boundary states are not protected against symmetry-breaking perturbations. Still, in what follows we show that boundary states of time-independent 1D chiral systems inherit a residual protection also against a large class of symmetry-breaking time-periodic perturbations. To be more specific, we show that boundary states of these systems show a resilience against perturbations of the form 
\begin{equation}
\label{eq:V}
V(t) = \sum_n V_n \cos(n \Omega t + \phi_n),
\end{equation}
 where $\forall n \in \mathbb{N} : \Gamma V_n \Gamma = \pm V_n$ ($\pm$ can depend on $n$) and $\phi_n \in \mathbb{R}$. Importantly, this class of perturbations which in general break chiral symmetry for all choices of reference time $t_0$, neither depends on the specifics of the model considered nor on any spatial fine-tuning.  To analytically uncover this surprising resilience of the boundary states we turn to Floquet perturbation theory and establish that  the expected leading-order quasienergy correction vanishes identically.

Analogous to conventional perturbation theory, Floquet perturbation theory allows us to estimate corrections to the eigenvalues (quasienergies) in powers of the strength of the time-periodic perturbation $V(t)$. Within this formalism the first- and second-order quasienergy corrections to any nondegenerate level are given~by  

 \begin{equation}
    \varepsilon^{1}_\psi = \frac{1}{T} \int^T_0\langle \psi^{0} (t) | V(t) | \psi^{0} (t)  \rangle dt,
 \label{eq:QS_1_conv}
\end{equation}
\begin{equation}
    \varepsilon^{2}_\psi = \sum_{ \beta \neq \psi}  \frac{\left| \frac{1}{T} \int^T_0 \langle \beta^{0} (t) | V(t) | \psi^{0}(t) \rangle dt \right |^{2}  }{\varepsilon^{0}_\psi - \varepsilon^{0}_\beta},
 \label{eq:QS_2_conv}
\end{equation}
where $| \psi^{0}(t) \rangle, | \beta^{0}(t) \rangle$ are unperturbed steady modes associated with quasienergies $\varepsilon^{0}_\psi$ and $\varepsilon^{0}_\beta$,  $V(t)$ is a time-periodic perturbation, and $T$ is a driving period. The sum runs over all steady modes $| \beta^{0}(t) \rangle$ differing from the mode under consideration $| \psi^{0}(t) \rangle$. The corrections above are given by the same expressions as in stationary perturbation theory but modified in accordance with the Floquet formalism, with matrix elements of operators being replaced by their time averages over the period $T$~\cite{Sambe}.

\begin{figure} \centering
    \includegraphics[width=8.0 cm, angle=0]{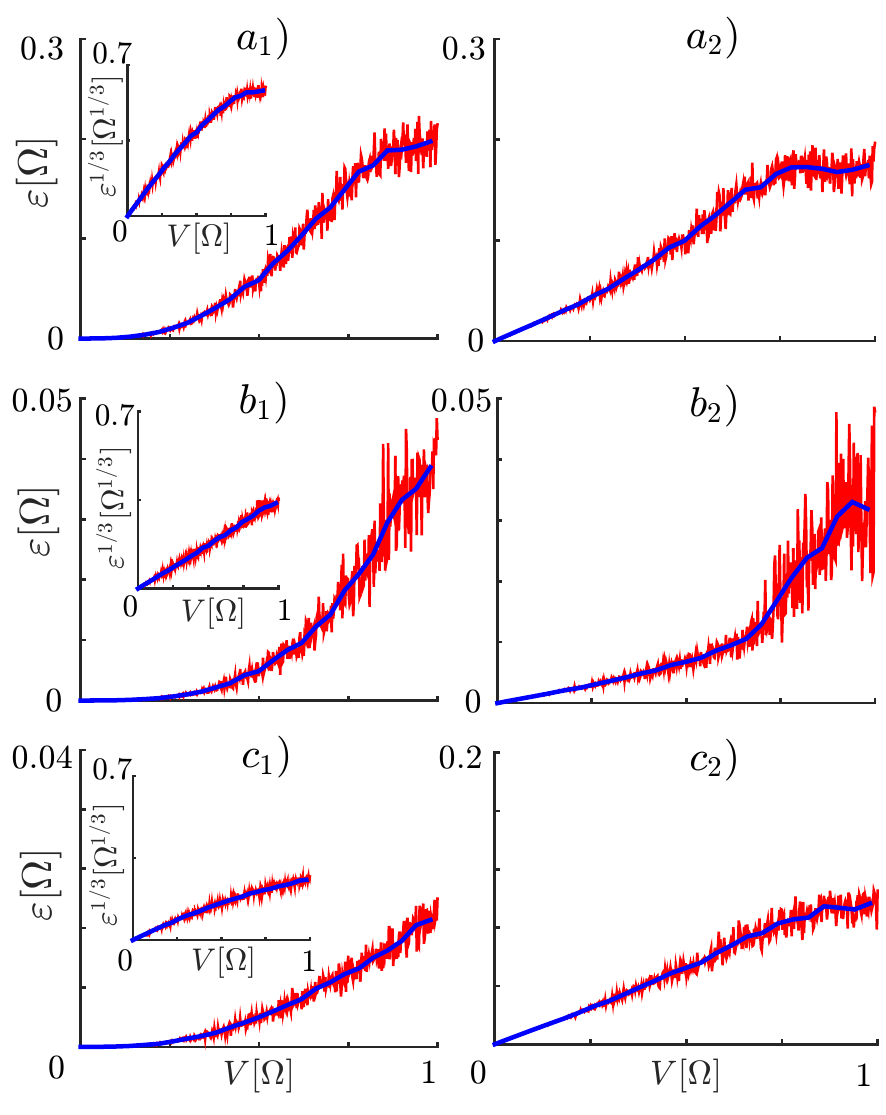}
        \caption{Scaling of maximum zero-quasienergy shifts with maximum disorder amplitude for time-independent (panels with subindex $1$) and harmonically driven (panels with subindex $2$) SSH chains with $\gamma_1 = 0.15 \,  \Omega$ and $\gamma_2 = 0.5 \,  \Omega$. The chains consist of 80 sites and in the driven case the harmonic modulation was fixed to $v(t) = 0.4 \cos(\Omega t)$. The added time-periodic boundary perturbation, extending over 20 sites from one of the edges, was implemented as a spatial disordering of two independent parameters: $a) \  \gamma_{1, j} \cos(\Omega t) + \Delta_j \cos(2 \Omega t)$; $b) \ \gamma_{2, j} \cos(\Omega t) + \gamma_{1, j} \sin(2 \Omega t)$; $c) \ \Delta_j \sin(\Omega t) + \gamma_{2, j} \sin(2 \Omega t)$ with $j = 1, 2, ... , 10$. In red color we plot the largest zero-quasienergy shift $\varepsilon[\Omega]$  maximized over 100 disorder realizations versus the largest allowed disorder site-amplitude, denoted by~$V [\Omega]$. The blue curve represents smoothed data obtained by replacing every 20 points by their average.}
     \label{fig_SSH_TIME_INDEP}
\end{figure}

Let us now consider a 1D topologically nontrivial system described by an unperturbed Hamiltonian $H_0(t)$. While we are here primarily interested in time-independent nonperturbed systems where $H_0(t) = H_0$, for now we keep the time argument in the Hamiltonian which allows us to discuss both cases on the same footing: driven {\em and} undriven nonperturbed systems. {
We assume that the system is chirally symmetric, implying that Eq.~(\ref{eq:SSH_chiral_symmetry_t}), $\Gamma H_0(t) \Gamma = - H_0( - t)$, is satisfied. For simplicity we suppress the reference time~$t_0$ in all formulas but keep in mind its presence whenever relevant. Equation~(\ref{eq:SSH_chiral_symmetry_t}) restricts the unperturbed steady modes to come in symmetry-bounded pairs: $| \beta^{0}(t) \rangle$ is a steady mode with quasienergy $\varepsilon^0_\beta$ if and only if $| \beta^{0}(- t) \rangle$ is also a steady mode but with quasienergy $-\varepsilon^0_\beta$. Also, it is assumed that in the thermodynamic limit the zero-quasienergy level is nondegenerate at each of the boundaries (with the zero-quasienergy boundary mode at the left edge having vanishingly small overlap with the zero-quasienergy boundary mode at the right edge), and therefore we may apply nondegenerate perturbation theory separately for each of the boundary modes. Without loss of generality we focus on the symmetry-protected mode satisfying $| \psi^{0}(t) \rangle = \Gamma | \psi^{0}(- t) \rangle$. This is the time-dependent analog of the relation $| \psi^{0}(0) \rangle = \Gamma | \psi^{0}(0) \rangle$ discussed in Appendix~B. Here, we focus on 1D chiral systems with only one localized state per edge but our approach can be straightforwardly generalized to the degenerate case, leading to the same result.

We are interested in the leading-order correction to the zero-quasienergy level $\varepsilon^{0}_\psi = 0$ associated with the state $| \psi^{0}(t) \rangle = \Gamma | \psi^{0}(-t) \rangle$ under influence of the time-periodic perturbations $V(t)$ introduced in Eq. (\ref{eq:V}). According to Eq.~(\ref{eq:QS_1_conv}), the first-order correction is generally nonzero for a driven state $| \psi^{0} (t) \rangle$; however, it does vanish in the case when $| \psi^{0} (t) \rangle = | \psi^{0} \rangle $ is a stationary state, i.e., an eigenstate of a time-independent system [$H_0(t)=H_0$]. This is so because of the integration over the period~$T$. This result is not surprising because the first-order correction represents energy-conserving transitions disallowed in time-independent systems by requiring the perturbative driving to have zero time average.

Given that the first-order correction vanishes identically when the unperturbed system is time-independent, we now consider the second-order correction~(\ref{eq:QS_2_conv}) for this case. The unperturbed modes $|\beta^0(t)\rangle$ are given now by $|\beta^0(t)\rangle = e^{i n \Omega t} |\beta^0\rangle$, with $n \in \mathbb{N}$ and $|\beta^0\rangle$ being eigenstates of the static Hamiltonian. By this, we can split the sum in (\ref{eq:QS_2_conv}) into a sum over quasienergy phases $e^{in\Omega t}$ and eigenstates $|\beta^0\rangle$, with $\varepsilon_{\beta}^0 \rightarrow \varepsilon_{\beta}^0 + n\Omega$. By using that $\varepsilon_{\psi}^0 = 0$, we thus obtain
\begin{align}
\begin{split}
    \varepsilon^{2}_\psi & = \sum_{\substack{\beta \neq \psi,  \, n }}  \frac{\iint \langle \beta^{0} | V(t) | \psi^{0} \rangle  \langle \psi^{0}| V(t^\prime) | \beta^{0} \rangle e^{i n \Omega(t-t^\prime)}} { - \varepsilon^{0}_\beta - n \Omega} \\
    & = \sum_{\substack{\beta \neq \psi, \, n}}  \frac{  \langle \beta^{0} | V^{(-n)} | \psi^{0} \rangle  \langle \psi^{0}| V^{(n)} | \beta^{0} \rangle }{ - \varepsilon^{0}_\beta - n \Omega}, \\
 \label{eq:QS_2_conv2_1}
\end{split} 
\end{align}
where $t$ and $t^\prime$ are both integrated over one period $T$ and then time-averaged (divided by $T$). Also, in the second line we have introduced the Fourier components $V^{(n)} \equiv \frac{1}{T} \int^T_0 e^{-i n \Omega t} V(t) dt$. It is easy to verify that these satisfy the relation $V^{(\pm n)} = e^{\pm i\phi_n} V_n /2$ in accordance with the assumed form of perturbations, Eq. (\ref{eq:V}). Together with the property  $\Gamma V_n \Gamma = \pm V_n$ and the chiral symmetry of the unperturbed Hamiltonian, implying that the unperturbed eigenmodes $|\beta^{0} \rangle$ with $\varepsilon^{0}_\beta$ always come paired to $ | \beta_\Gamma^{0} \rangle \equiv \Gamma | \beta^{0} \rangle $ with $\varepsilon^{0}_{\beta_\Gamma} \equiv - \varepsilon^{0}_{\beta}$, this relation allows us to derive 
\begin{align}
\begin{split}
    &\varepsilon^{2}_\psi  = \sum_{\substack{\beta \neq \psi }} \sum_n \frac{  \langle \beta^{0} | V^{(-n)} | \psi^{0} \rangle  \langle \psi^{0}| V^{(n)} | \beta^{0} \rangle }{ - \varepsilon^{0}_\beta - n \Omega}  \\
    &= \sum_{\substack{\beta \neq \psi  }}  \sum_n \frac{  \langle \beta^{0} |\Gamma \Gamma V^{(-n)} \Gamma \Gamma | \psi^{0} \rangle  \langle \psi^{0} |\Gamma \Gamma  V^{(n)}  \Gamma \Gamma | \beta^{0} \rangle }{ - \varepsilon^{0}_\beta -n \Omega}\\ 
    & = \sum_{\substack{\beta \neq \psi }} \sum_n \frac{  \langle \beta_\Gamma^{0} | V^{(-n)} | \psi^{0} \rangle  \langle \psi^{0}| V^{(n)} | \beta_\Gamma^{0} \rangle }{ \varepsilon^{0}_{\beta_\Gamma} + n \Omega} \\ 
    &= 0.  \\
 \label{eq:QS_2_conv2_2}
\end{split} 
\end{align}

The correction $\varepsilon^{2}_\psi$ vanishes because the first and last expressions in Eq.~(\ref{eq:QS_2_conv2_2}) are the same but with opposite signs (only the terms in the summations are ordered differently). As follows from the derivation, the vanishing of the second-order correction for this class of perturbations crucially hinges on the chiral symmetry of the unperturbed Hamiltonian.

In Fig.~5 we numerically verify the scaling of the quasienergy shifts when perturbed by various driven disorders, again employing the SSH model. To test our prediction we choose to disorder two independent parameters and drive the corresponding perturbations periodically in time by using a superposition of a first and second harmonic. Such perturbations may be less accessible experimentally but represent well the class of perturbations assumed in the calculation above. In agreement with our perturbative prediction, in the undriven case the leading-order scaling is indeed only cubic in the disorder strength (see Fig. 5). The deviations from cubic scaling come from subleading terms, which, as expected, grow with the disorder strength. For comparison, we have also considered harmonically driven chains under analogous disordering perturbations and found that the leading-order scaling is here linear, as anticipated.

\subsection{A reduced effect of symmetry-breaking perturbations outside the perturbative regime} 

\begin{figure*}
  \includegraphics[width=\textwidth]{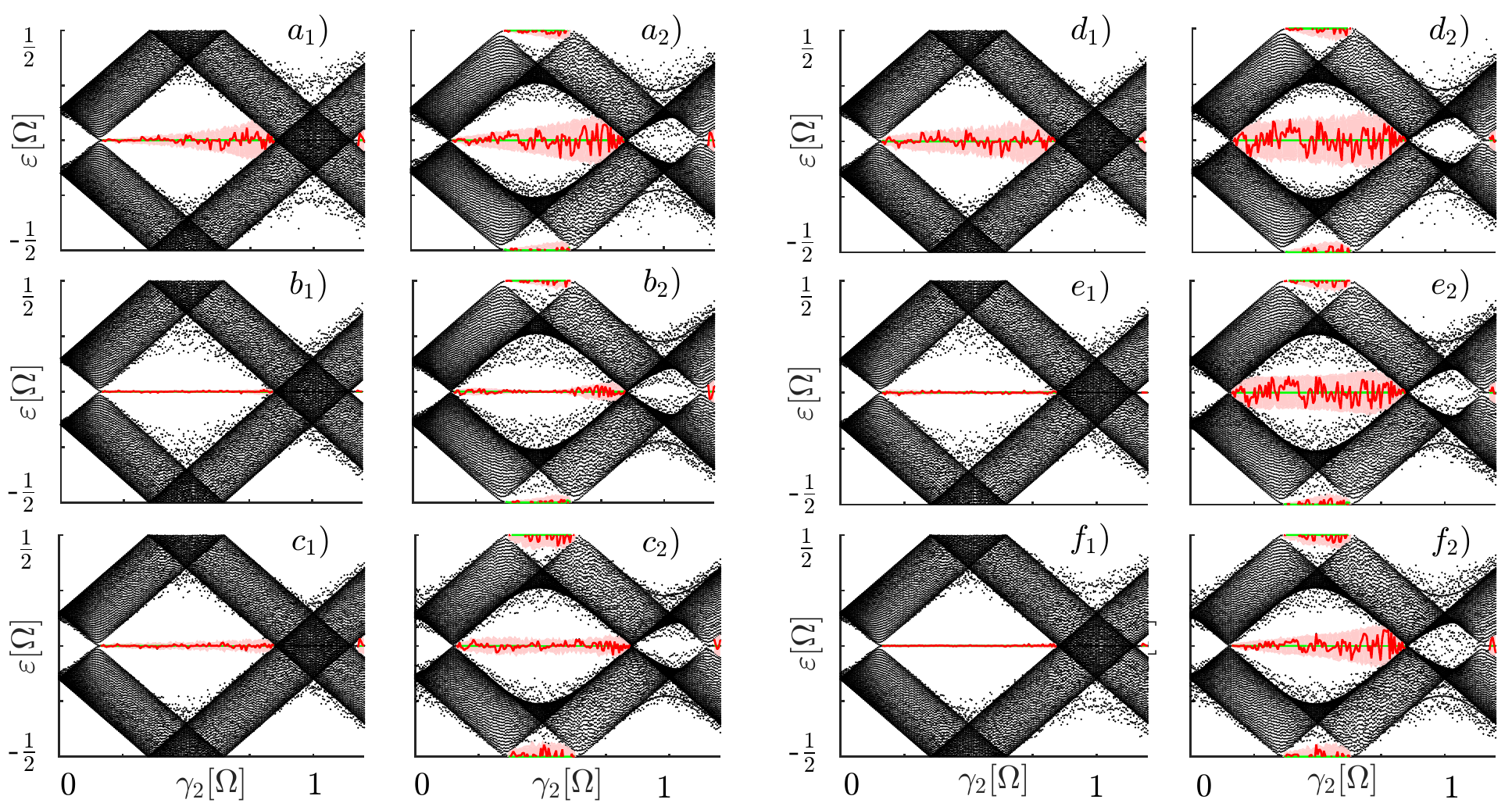}
  \caption{Quasienergy spectra obtained from disordering the first 20 sites of 80-site SSH chains with unperturbed $\gamma_1 = 0.15 \, \Omega$ and $\gamma_2$ swapped from $0$ to $1.2 \,  \Omega$. The levels for bulk states, perturbed and unperturbed edge states are displayed in black, red, and green, respectively. The panels having subindex $1$ or $2$ correspond to the time-independent or harmonically driven [with $v(t) = 0.4 \cos(\Omega t)$] cases, respectively. The disorder is here added as a time-periodic spatial disordering of the amplitudes for hopping ($\sim \gamma_1, \gamma_2$) and/or a staggered chemical potential ($\sim \Delta$): $a) \  \gamma_{1, j} \cos(\Omega t) + \gamma_{2,j} \sin(2 \Omega t)$; $b) \  \gamma_{2, j} \cos(\Omega t) + \gamma_{1,j} \sin(2 \Omega t)$; $c) \  \Delta_{j} \sin(\Omega t) + \gamma_{1,j} \sin(2 \Omega t)$; $d) \  \gamma_{1,j} \cos(\Omega t) + \Delta_{j} \cos(2 \Omega t)$; $e) \  \gamma_{2,j} \cos(\Omega t) + \Delta_{j} \cos(2 \Omega t)$;   $e) \  \Delta_{j} \sin(\Omega t) + \gamma_{2,j} \sin(2 \Omega t)$ with $j = 1,..., 10$. The maximum disorder strength was taken to be $0.5 \, \Omega$ in each of the disordering parameters. The pink areas define the regions to which the shifted midgap quasienergies were found to be confined after collecting data from $100$ distinct disorder realizations.}
\end{figure*}

The analysis above indicates that the quasienergy shifts in chiral time-independent systems are typically small in the perturbative regime for all time-periodic 
disorders with vanishing static components. This is so since at least the first-order correction vanishes in this case. Here we present a qualitative argument why one should 
expect a reduced effect also for strong disorders. 

 We recall that time-independent chiral systems possess a chiral symmetry for any choice of reference time~$t_0$. Decomposing a generic time-periodic perturbation as $V=V_{\text{S}} + V_{\text{nS}}$, where $V_{\text{S}}\, (V_{\text{nS}})$ preserves (breaks) chiral symmetry, $t_0$ can then be chosen so as to minimize $V_{\text{nS}}$, denote it by $V^{\text{min}}_{\text{nS}}$. In general, the  picked reference time changes as we vary the time-dependence of $V.$ Inasmuch as a quasienergy shift is insensitive to the choice of reference time, it then has to remain small whenever $V_{\text{nS}}^{\text{min}}$ is small (since the symmetry-preserving component by itself does not affect the midgap quasienergy). It~is essential to point out that the minimized symmetry-breaking part $V_{\text{nS}}^{\text{min}}$~can be small even when the total perturbation $V$ is large. By this, one expects that the redundancy of possible symmetry-respecting reference times intrinsically bolsters an enhanced resilience of the boundary states compared with systems without this degree of freedom. 
 This makes static chiral systems under time-periodic perturbations special, different from chiral Floquet systems (with the symmetry-respecting reference time fixed by the driving in the bulk) and time-independent systems under static perturbations (where the symmetry-breaking component of the perturbation does not at all depend on the reference time).
 
 To  assess  this  statement, we present numerical results performed on the strongly disordered time-independent and harmonically-driven SSH models. In all cases considered the quasienergy shifts in the time-independent SSH chains are found to be suppressed compared with those in the harmonically driven case (see Fig. 6). Notably, in several of the cases the suppression is dramatic, implying extremely resilient boundary levels.

\subsection{Time-independent SSH model under time-periodic disorder in the chemical potential}

In the numerical studies of the time-independent SSH model we have so far examined two types of boundary perturbations: time-periodic spatial disordering of the hopping amplitudes $(\sim \gamma_1, \gamma_2)$ and of an added staggered chemical potential term $(\sim\Delta)$. Both of these perturbations are probably difficult to implement in an experimental setup. A boundary perturbation which is expected to have greater potential to be realized in an experiment is that of a time-periodic spatially disordered chemical-potential term.  Adding it to the time-independent SSH model, we have the Hamiltonian
\begin{align}
\begin{split}
H(t)  = & - \sum_{j} \left( \gamma_1  c^\dagger_{A,j} c^{\phantom\dagger}_{B,j} 
+ \gamma_2 c^\dagger_{B,j-1} c^{\phantom\dagger}_{A,j} + \text{H.c.} \right ) \\
& +\sum_{\sigma, j}  \mu_{\sigma, j}(t) c^\dagger_{\sigma,j} c^{\phantom\dagger}_{\sigma,j}.
\label{eq:SSH_harmonic_chem}
\end{split} 
\end{align}

Here the first sum represents the time-independent SSH chain, while the second term, with $\sigma = A,B$, is the added chemical potential, disordered in a boundary region by allowing the time-periodic amplitudes $\mu_{\sigma, j}(t)$ to vary randomly for lattice sites $j$ in some neighborhood of one of the edges of the chain. For the rest of the chain the chemical potential is set to zero. It is interesting to note that the disorder in the chemical potential can be eliminated at the cost of introducing a time-periodic disordering of the hopping amplitudes (see Appendix~C). It follows that a perturbation from a time-periodic disordered chemical-potential term results in the same physics as we uncovered above when considering time-periodic disordering of the hopping amplitudes. 

A perturbation in the chemical potential commutes with $\Gamma$ and therefore it has to be {\it odd} in time for some reference time $t_0$ in order to respect chiral symmetry and by this robustness of the SPT states. Accordingly, whereas a static disorder in a chemical potential kicks the symmetry-protected boundary states away from zero quasienergy, any harmonically driven disorder will preserve a chiral symmetry at a certain reference time and have no effect on the corresponding quasienergies. Things change if adding a second driving harmonic. If it is out of phase with the first harmonic then all chiral symmetries get broken and the midgap states are no longer protected. The numerical data in Fig.~7, obtained for an SSH chain with 80 sites, fully support this picture. 

\begin{figure} \centering
    \includegraphics[width=8.0 cm]{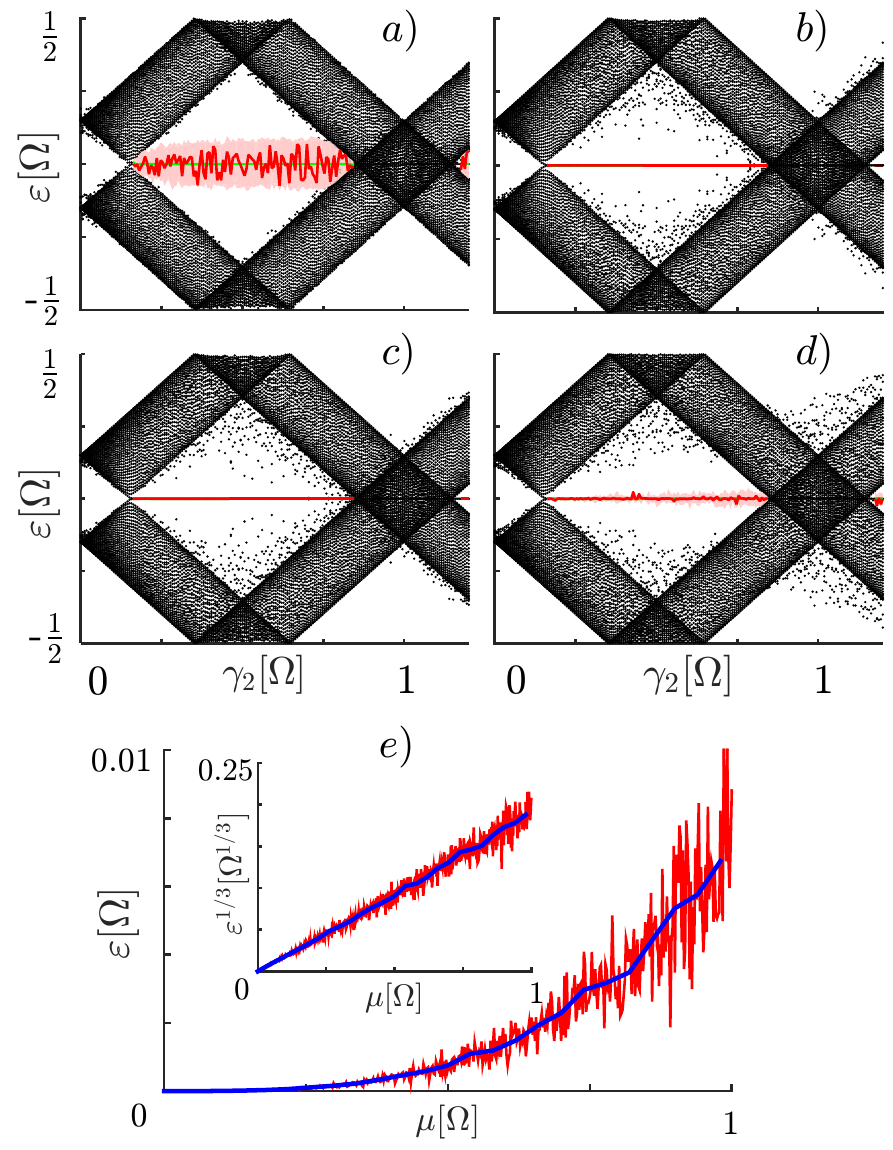}
        \caption{Numerical data obtained from time-independent SSH chains [$v(t) = 0$ in Eq. (1)] with 80 sites subject to static and time-periodic boundary perturbations, here realized as a spatial disordering of an added chemical-potential term ($\sim \mu$) and extending over 20 sites from one of the boundaries. Panels $a) - d)$ illustrate quasienergy spectra for the chains with unperturbed hopping amplitudes $\gamma_1 =  0.15 \, \Omega$ and $\gamma_2$ varying from $0$ to $1.2 \, \Omega$. The boundary perturbations were considered to be of the form $a) \ \mu_{\sigma, j}$ (static disorder),  $b) \  \mu_{\sigma, j} \sin(\Omega t)$, $c) \  \mu_{\sigma, j} \cos(\Omega t) $, and $d) \ \mu_{\sigma, j} [ \sin(\Omega t) +  \cos(\Omega t)]$ with $\sigma = A, B$ and $j = 1, ...,  10$. The disordered site-amplitudes $\mu_{\sigma, j}$ vary randomly in the interval $[-0.1 \, \Omega, 0.1 \, \Omega]$ ($[-1.0 \, \Omega, 1.0 \, \Omega]$) in the static (time-dependent) cases. The levels for the bulk states and perturbed and unperturbed edge states are colored black, red, and green, respectively. $100$ disordering configurations were checked, with the midgap quasienergies always found to be within the corresponding pink areas. In panel $e)$ we present the scaling of the midgap quasienergy shifts in the time-independent SSH chains ($\gamma_1 =  0.15 \, \Omega$ and $\gamma_2 = 0.5 \, \Omega$) perturbed by time-periodic boundary disorder in chemical potential $\sim \mu [ \sin(\Omega t) +  \cos(\Omega t)]$. Here the largest midgap quasienergy shift $\varepsilon[\Omega]$ maximized over 500 disorder realizations is plotted in red versus the upper limit of the on-site perturbation amplitude in the chemical potential, denoted by~$\mu[\Omega]$. The blue curve represents smoothed data obtained by replacing every 20 points by their average.}
     \label{Perturbation}
\end{figure}

In  Fig.~\ref{Perturbation} we also numerically verify the scaling of the quasienergy shifts when perturbed by a disordered chemical potential term driven by $\sin(\Omega t) + \cos(2\Omega t)$, a perturbation belonging to the class defined by Eq.~(4).  In good agreement  with  theory, the obtained leading-order scaling is only cubic in the disorder strength. As in Fig.~5 the deviations from cubic scaling are result of subleading terms, which become larger with the disorder strength.

\subsection{Experimental test}

Most ingredients needed to experimentally test our predictions have already been realized in optically trapped cold atoms \cite{Bloch, Bloch2, Jan}. Starting with the realization of the time-independent (undriven) SSH model, it has recently been simulated by trapping atoms in a specially designed 1D optical lattice~\cite{Atala, CA_SSH_states}. As for imaging states in cold-atomic setups, there now exists a number of proposals how to do this~\cite{Bloch, Bakr, Sherson, Endres, Islam}, and quite recently SSH boundary modes were observed in an experiment using optical real-space imaging~\cite{CA_SSH_states}.  Time-dependent driving of optical lattices has also been carried out in the laboratory \cite{CA_driven, CA_driven2}. Moreover, an experimental design for realizing the harmonically driven SSH model using cold atoms trapped in a dynamic optical lattice has recently
been put forward \cite{Lago}, awaiting practical implementation. Finally, efficient means of producing and studying static \cite{CA_static_disorder, CA_static_disorder2, CA_static_disorder3} and time-periodically driven \cite{CA_driven_disorder} disorder in cold-atomic simulations are by now routine. In light of these advances, an experimental test of our results appears fully viable. 

\section{Summary} 

 In this article, using Floquet theory and with the SSH model as a touchstone, we have carried out a systematic study of how the chiral symmetry protection in a topological insulator extends to states subject to time-periodic perturbations $-$ with the topological phase realized either as a Floquet topological phase or as a conventional time-independent one. Intriguingly, in a time-independent topological phase we found that the edge states exhibit an unexpected resilience against a large class of symmetry-breaking time-periodic perturbations as a result of the very structure of the unperturbed chiral-symmetric spectrum. This outstanding feature should be possible to test in an experiment with cold atoms. The extension of our analysis to other classes of symmetry-protected topological phases remains an interesting open problem.

\section*{ACKNOWLEDGMENTS}

 It is a pleasure to thank Jan Budich and Wen-Long You for useful discussions. We are also grateful to an anonymous referee for prodding us to undertake a perturbative analysis of the robustness of the boundary states. This work was supported by the Swedish Research Council through Grant No. 621-2014-5972. \\ 

\appendix 

\section{Conservation of Chiral Symmetry in Floquet Systems}

Here we prove that in order for chiral symmetry to be preserved in a Floquet system it is sufficient that the relation $\Gamma H(t) \Gamma = - H(-t)$ is satisfied, where $H(t)$ is a time-dependent Hamiltonian with period $T$ and $\Gamma$ is the corresponding unitary operator dictated by chiral symmetry. 

We start by defining evolution operators for consequent halves of the period $T$, $F\equiv U(0,\frac{T}{2})$ and $G\equiv U(\frac{T}{2}, T)$. Then, directly from the definition
\begin{align}
\begin{split}
&F \equiv \sum_n (-i)^n \int_{0}^{\frac{T}{2}} dt_1 \, ...  \int_{0}^{t_{n-1}} dt_n \, H(t_1) \, ... \, H(t_n),  \\
&G  \equiv \sum_n (-i)^n \int_{\frac{T}{2}}^{T} dt_1  \, ...  \int_{\frac{T}{2}}^{t_{n-1}}  dt_n \, H(t_1)  \, ... \, H(t_n). \\
\end{split}
\label{eq:F_G}
\end{align}

The Hamiltonian~$H(t)$ is periodic in time and therefore $G$ also equals $U(-\frac{T}{2}, 0)$. We then use the substitution $\forall i \in \mathbb{N}: \, \tau_i = -t_i$ and the condition  $\Gamma H(t) \Gamma = - H (-t)$ to obtain a relation between $F$ and~$G$, 
\begin{align}
\begin{split}
&F = \sum_n (i)^n \int_0^{-\frac{T}{2}} d\tau_1 \,... \, \int_0^{\tau_{n-1}} d\tau_n H(-\tau_1) \, ... \,      H(- \tau_n) \\  
& = \sum_n (-i)^{n} \int_0^{-\frac{T}{2}} d\tau_1 \, ... \, \int_0^{\tau_{n-1}} d\tau_n \Gamma H(\tau_1) \Gamma \, ... \, \Gamma H( \tau_n) \Gamma \\
& = \Gamma U(0, - \frac{T}{2}) \Gamma  = \Gamma G^\dagger \Gamma. 
\end{split}
\label{eq:the_relation}
\end{align}
The chiral symmetry condition  $\Gamma U(0,T) \Gamma = U^{-1}(0,T)$~\cite{Asboth, Asboth2} then follows immediately from $U(0,T) = F G = \Gamma G^\dagger \Gamma G$.

\section{Protection by Chiral Symmetry}

The symmetry protection of the edge states originates from the necessity of the eigenstates to come in pairs whenever the relation $\Gamma U(0,T)\Gamma = U^{-1}(0,T) $ is fulfilled. More precisely, the state $|u(0)\rangle$ is an eigenstate of $U(0,T)$ with eigenvalue $\exp(-i\varepsilon T)$ if and only if the state  $\Gamma|u(0)\rangle$ is also an eigenstate but with eigenvalue $\exp(i\varepsilon T)$. By unitarity of $U(0,T)$, any eigenvectors with different eigenvalues are orthogonal and therefore each pair of the symmetry-controlled states correspond to two distinct steady states except when the quasienergy $\varepsilon$ is $0$ or $\Omega/2$. For these two quasienergies the eigenvectors $|u(0) \rangle$ and its complement $\Gamma|u(0)\rangle$ have the same eigenvalues and therefore can be combined to form the states $P_A |u(0) \rangle$ and $P_B |u(0) \rangle$, where the orthogonal sublattice-projecting operators are defined as $P_A = (1 + \Gamma)/2$ and $P_B = (1 - \Gamma)/2$. Thus, we can always consider the $0$ and $\Omega/2$ quasienergy modes to have support on only one sublattice, and, as a consequence, being identical to their chiral symmetry partners. It follows that in the thermodynamic limit any symmetry-preserving perturbation of $U(0,T)$ which leaves the spectral gaps open, cannot change the difference $N_{A, \varepsilon} - N_{B, \varepsilon}$, where $N_{A, \varepsilon}$ and $N_{B, \varepsilon}$ are the number of states, localized at one of the edges, at  quasienergy $\varepsilon =0$ or $\Omega/2$ with support on sublattices $A$ and $B$, respectively. This is simply because, after any such perturbation, the states may come or leave the quasienerergies $0$ or $\Omega/2$ only in pairs (as dictated by the chiral symmetry), and therefore in the thermodynamic limit the difference $N_{A, \varepsilon} - N_{B, \varepsilon}$ cannot change. These differences in edge-state numbers are exactly the topological invariants $\nu_0$ and $\nu_\pi$ calculated in Fig.~2, using a sublattice polarization argument as elaborated in Refs.~\cite{Asboth, Asboth2}. To summarize, any gap-preserving perturbation of $U(0,T)$ respecting the chiral symmetry cannot change the number of symmetry-protected states at quasienergies $0$ and~$\Omega/2$. This implies that any periodically driven boundary perturbation that does not break the chiral symmetry of $U(0,T)$ leaves the eigenvalues of the protected edge states unaffected. 

\section{Gauging out the Perturbation of the Chemical Potential}

The time-independent SSH model disordered in the chemical potential is described by Eq.~(9). In the following it is shown that we can drop the disordering term in Eq.~(9) at the cost of introducing some extra disorder in the hopping amplitudes. For this purpose we define the unitary operator 
\begin{align}
\begin{split}
U(t)  = e^{- i  \sum_{\sigma,j} M_{\sigma,j}(t)  c^\dagger_{\sigma,j} c^{\phantom\dagger}_{\sigma,j} },
\label{eq:unitary_transformation}
\end{split} 
\end{align}
where $\frac{d}{dt} M_{\sigma,j}(t) = \mu_{\sigma, j}(t)$. By working out the commutation relations between $U(t)$ and the disordered SSH Hamiltonian $H(t)$ given in Eq.~(9), we arrive at the following transformed Hamiltonian $H^\prime  \equiv U^\dagger  H U - i U^\dagger \frac{d}{dt} U$, with
\begin{align}
\begin{split}
& H^\prime(t)   = - \sum_{j} \left( e^{i(M_{A,j}(t) - M_{B,j}(t))} \gamma_1  c^\dagger_{A,j} c^{\phantom\dagger}_{B,j} \right.\\
& \left. + \ e^{i(M_{B,j-1}(t) - M_{A,j}(t))}\gamma_2  c^\dagger_{B,j-1} c^{\phantom\dagger}_{A,j} + \text{H.c.} \right). \\
\label{eq:unitary_transformation_eq}
\end{split} 
\end{align}

The transformed Hamiltonian $H^\prime(t)$ describes the same system as $H(t)$ and it has the form of an SSH Hamiltonian but with disordered hopping amplitudes. Moreover, if all $M_{\sigma,j}(t)$ are periodic in time then $H^\prime(t)$ can also be handled within the Floquet formalism, yielding the same quasienergy spectrum as the original system. In order for $M_{\sigma,j}(t)$ to be periodic in time, all the time-periodic disorder amplitudes $\mu_{\sigma, j}(t)$ must have vanishing zeroth components in their Fourier series decompositions (the static components). This calculation explicitly reveals the equivalence of time-periodic disorders in the chemical potential and a certain type of disorders in the hopping amplitudes.

\end{document}